\documentclass[10pt, conference, compsocconf]{IEEEtran}
\ifCLASSINFOpdf
   \usepackage[pdftex]{graphicx}
   \graphicspath{{../pdf/}{../jpeg/}}
   \DeclareGraphicsExtensions{.pdf,.jpeg,.png}
\else
   \usepackage[dvips]{graphicx}
   \graphicspath{{../eps/}}
   \DeclareGraphicsExtensions{.eps}
\fi
\begin{document}
\title{Semi-metric networks for recommender systems}


\author{\IEEEauthorblockN{Tiago Simas}
\IEEEauthorblockA{Cognitive Science Program\\
Indiana University, Bloomington, IN 47406, USA\\
Email: tdesimas@indiana.edu}
\and
\IEEEauthorblockN{Luis M. Rocha}
\IEEEauthorblockA{Center for Complex Networks and Systems\\
School of Informatics \& Computing\\
Indiana University, Bloomington, IN 47406, USA \\
Email: rocha@indiana.edu}
}


%Telephone: (800) 555--1212, Fax: (888) 555--1212}


\maketitle

\begin{abstract}
Weighted graphs obtained from co-occurrence in user-item relations lead to non-metric topologies. We use this semi-metric behavior to issue recommendations, and discuss its relationship to transitive closure on fuzzy graphs. Finally, we test the performance of this method against other item- and user-based recommender systems on the \emph{Movielens} benchmark. We show that including highly semi-metric edges in our recommendation algorithms leads to better recommendations. %The abstract goes here. DO NOT USE SPECIAL CHARACTERS, SYMBOLS, OR MATH IN YOUR TITLE OR ABSTRACT.
\end{abstract}


% \fi
\IEEEpeerreviewmaketitle

\newtheorem{Algorithm}{Algorithm}

\section{Introduction: Recommendation as Prediction}
The identification of association or correlation between time events is important for many systems, such as: recommender systems, social behavior, functional brain interaction, event-detection, financial forecasting, and many more.
Recommender systems are a good example of prediction, since the goal is to recommend the items users may be interested in the future, given information about how they accessed or purchased items in the past \cite{herlocker99}.
Recently, there has been much interest in the analysis of complex networks \cite{pastor_vesp}---extracted from large collections of textual documents and user access patterns---to predict social behavior including online behavior \cite{IPAM_social_data_mining}.
% \cite{Conover2010predicting}.
%
In previous work, we developed complex network methods to uncover clusters in non-metric network topologies that arise in weighted graphs obtained from real-world data (e.g. via co-occurrence statistics, see below). Our clustering methodology, which is equivalent to what has become known more recently as link communities\cite{Ahn2010b}, has been applied to social networks,
%
%\cite{Rocha2002_terr},
%
word networks, scientific journal networks, etc [e.g.\cite{rocha05mylib,abihaidar_GB08}].
%
%and citation networks \cite{rocha05mylib}\cite{verspoor05BMC,abihaidar_GB08}\cite{Kolchinsky2010a}.

Of particular interest to prediction in recommendation, we have developed measures to extract the graph edges which most violate the triangle inequality: \emph{semi-metric associations} (see below). Our working hypothesis is that strong semi-metric associations can be used to identify items with a higher probability of co-occurring in the future, as well the dynamics of such networks in general \cite{Rocha2002}.
%
%This methodology has been applied to recommender systems for the digital library at the \emph{Los Alamos National Laboratory}\cite{rocha05mylib}, the \texttt{givealink.org} project \cite{givealink2006}, networks of felons obtained from intelligence records \cite{Rocha2002_terr}, etc.
%
This methodology has been applied to recommender systems for the digital library at the \emph{Los Alamos National Laboratory}, the \texttt{givealink.org} project, networks of felons obtained from intelligence records, etc.
The performance of this approach was assessed using expert evaluations \cite{rocha05mylib}. While this performance assessment showed that recommendations issued on the basis of semi-metric behavior were relevant to users, one has to worry about the subjectivity of human experts. Moreover, it did not allow us to conclude about the ability of semi-metric associations to predict future user choices in recommender systems.
To address these concerns, here we use the \emph{MovieLens} benchmark\footnote{http://movilens.umn.edu}.
%
%, to test the accuracy of the recommendations, and compare with some previous work done on the same data.
%
The advantage of using this benchmark is that it has been widely used to assess various recommender systems in the literature. The disadvantage is that the results are specific to the Movilens database on the topic of movies preferences only. There are other datasets, such as the one provided by Netflix\footnote{www.netflix.com}, which we will address in future work. Here, we simply want to establish, without expert subjectivity, that semi-metric behavior can be useful to predict future user behavior and thus issue quality recommendations; to achieve that goal, as we show below, the MovieLens benchmark is sufficient.

\section{Background}
\subsection{Knowledge extraction in Proximity Graphs}

Our approach starts with probabilistic proximity measure computed
from binary relations between any two sets of items (e.g. keywords-documents or items-users).
This measure is a natural weighted
extension \cite{Rocha1999}
% \cite{rochabollen2001}
\cite{Popescu2006} of the Jaccard similarity measure \cite{Grefenstette94}, which has been used extensively in computational intelligence \cite{nakamura} %\cite{rocha02Terr}  \cite{Rocha2002} \cite{Rocha2005}
\cite{rocha01talkmine}.
Given a generic binary relation $R$ between sets $X$
(of $n$ elements $x$) and $Y$ (of $m$ elements $y$), we extract two
complementary \emph{proximity graphs}: $XYP$ and $YXP$.

\begin{equation}
\label{proximity_measures}
xyp_{i,j}=\frac{\displaystyle\sum_{k=1}^m (r_{ik} \wedge
r_{kj})}{\displaystyle\sum_{k=1}^m (r_{ik} \vee r_{kj})};
yxp_{i,j}=\frac{\displaystyle\sum_{k=1}^n (r_{ki} \wedge
r_{kj})}{\displaystyle\sum_{k=1}^n (r_{ki} \vee r_{kj})}
\end{equation}

These measures equate proximity with co-occurrence. $xyp(x_i, x_j)$ is the probability that
both $x_i$ and $x_j$ are related (co-occur) via $R$ to the same elements $y \in
Y$ (and only those)---and vice-versa for $yxp$.
Below, when we refer to a proximity graph $P$, we mean a graph obtained via formula \ref{proximity_measures}.
Other co-occurrence measures can be used to capture a degree of
proximity between elements of two sets in a binary
relation. In information retrieval, it is common to use the cosine \cite{BaezaYates99},
Euclidean \cite{Strehl} and even mutual information measures
\cite{turney01}. For characterizing closeness in relations, we
prefer our weighted Jaccard proximity measure because it possesses
several desirable characteristics. %Mutual information-based measures
%are not symmetric, therefore they are neither proximity nor
%similarity measures as defined above.
The Euclidean measure is a
similarity measure (it is transitive), but it generates non-sparse matrices,
since all finite elements of the relation $R$ lead to similarity
greater than zero. This makes it impractical for very large data
sets. The cosine proximity measure (which is typically not transitive) is scale-invariant which makes it very
appealing for text documents of varying size, but may be problematic in other
domains. The weighted Jaccard measure has aspects of both the
Euclidean and the cosine measures \cite{Strehl}, and leads to
sparse matrices.
%
%We use our weighted extension of
%Jaccard proximity measure (eq. \ref{proximity_measures}) in several applications. However, all of the theoretical work we propose below applies to any proximity graph (as defined above), independently of the measure used to obtain it from specific data sets.
%\subsection{Knowledge extraction from network modularity}

Proximity graphs can be seen as
\emph{associative knowledge networks} that represent how often items
co-occur in a large set of documents \cite{Rocha2002, rocha03NATO}.
The assumption is that items
that frequently co-occur, are associated with a common concept
understood by the community of users and writers of the documents.
Notice that a graph of co-occurrence proximity allows us to capture
network associations rather than just pair-wise co-occurrence. In
other words, we expect concepts or themes to be organized in more
interconnected sub-graphs, or clusters of items in the proximity
networks.
Indeed, we have successfully used the modularity of proximity networks in
several knowledge extraction and literature mining applications, from recommender systems \cite{rocha05mylib} to biomedical text mining \cite{verspoor05BMC,abihaidar_GB08}.
More recently, modularity-detection in proximity graph has been rediscovered in the literature as the idea of \emph{link communities} \cite{Ahn2010b}, which applies the Jaccard similarity measure to graphs prior to identification of clusters.

\subsection{Transitive and Distance Closure}
Proximity graphs are reflexive and symmetric fuzzy graphs.
We can perform a transitive closure of these graphs using the composition of their connectivity matrices,
which is done in much the same way as the algebraic
composition of matrices, except that multiplication and summation
are substituted by generalized fuzzy logic conjunctions ($\wedge$) and disjunction ($\vee$),  more generally known as T-Norms and T-Conorms
respectively \cite{klir-2001}.
%
%
%\begin{equation}
%\label{proximity_measures}
%xyp_{i,j}=\frac{\displaystyle\sum_{k=1}^m (r_{ik} \wedge
%r_{kj})}{\displaystyle\sum_{k=1}^m (r_{ik} \vee r_{kj})};
%yxp_{i,j}=\frac{\displaystyle\sum_{k=1}^n (r_{ki} \wedge
%r_{kj})}{\displaystyle\sum_{k=1}^n (r_{ki} \vee r_{kj})}
%\end{equation}

$$P \circ P = \displaystyle\bigvee_{k} \bigwedge (p_{ik},p_{kj})=p'_{ij}$$
\noindent where $P$ denotes a proximity graph, and $p_{i,j} \in [0,1]$ the entries of its connectivity matrix.
The most commonly used operations are $\wedge =$\emph{minimum} (conjunction) and
$\vee =$ \emph{maximum} disjunction. But there are many large classes of such functions available \cite{klir-2001}.
The \emph{transitive closure} $P^\infty$ of a proximity graph $P$ is obtained via the following algorithm\cite{klir-2001}:
\begin{enumerate}
  \item $P' = P \circ P$
  \item If $P'\neq P$, make $P = P'$ and go back to step 1.
  \item  Stop: $P^\infty = P'$
\end{enumerate}
The transitive closure of $P$ yields a similarity graph.

Instead of a proximity graph, it is often useful to work with a distance graph $D$, where $d_{i,j} \in [0,\infty]$, $d_{i,i}=0$, $d_{i,j}=d_{j,i}$. In this case, instead of proximity/similarity, edge weights denote dissimilarity represented with the very intuitive notion of distance.
Similarly, we can compute a \emph{distance closure}, $D^\infty$ to compute the smallest possible distance between vertices. This is done in exactly the same way as the transitive closure, except that matrix composition becomes  $D \circ D = f_{k}(g(d_{ik},d_{kj}))=d'_{ij}$, for a pair of monotonic functions $f,g$, which we have referred to elsewhere as TD-Conorms and TD-Norms \cite{simas2011}.
A special case of distance closure is the \emph{metric closure}, where $f(x,y)=min(x,y)$ and $g(x,y)=x+y$. This type of closure computes the shortest path between all edges in $D$ --- it is thus equivalent to the All Pairs Shortest Paths (APSP) algorithm \cite{Zwick}.

We can define an isomorphism between the two types of graphs and closures, but only by using a non-linear map $\varphi$, since proximity edges are constrained to $[0,1]$, while distance edges to $[0,+\infty]$ \cite{simas2011}. To establish an isomorphism (for graphs $P$ and $D$ to commute), we must guarantee:

\[\forall i,j \in P:\mathop{f}\limits_{k} \{g(\varphi (p_{i,k}),\varphi (p_{k,j})\}=\varphi(\mathop{\lor}\limits_{k}\{ \land (p_{i,k},p_{k,j})\})\]

\noindent which leads to the equations that allow us to define the constraints of each operation:
%\\
%\[g(d_{i,k},d_{k,j})=\varphi(\land(\varphi^{-1}(d_{i,k}),\varphi^{-1}(d_{k,j})))\]
%\[f(d_{i,k},d_{k,j}) \equiv \varphi (\lor( \varphi^{-1}(d_{i,k}),\varphi^{-1}(d_{k,j})))\]
%\[\lor(p_{i,k},p_{k,j})=\varphi^{-1}(f(\varphi(p_{i,k}),\varphi(p_{k,j})))\]  \[\land(p_{i,k},p_{k,j})=\varphi^{-1}(g(\varphi(p_{i,k}),\varphi(p_{k,j})))\]

\begin{eqnarray}
\label{isomorphism_formulae}
g(d_{i,k},d_{k,j}) &=& \varphi(\land(\varphi^{-1}(d_{i,k}),\varphi^{-1}(d_{k,j}))) \nonumber \\
f(d_{i,k},d_{k,j}) &=& \varphi (\lor( \varphi^{-1}(d_{i,k}),\varphi^{-1}(d_{k,j})))\nonumber \\
\lor(p_{i,k},p_{k,j}) &=& \varphi^{-1}(f(\varphi(p_{i,k}),\varphi(p_{k,j}))) \nonumber \\
\land(p_{i,k},p_{k,j}) &=& \varphi^{-1}(g(\varphi(p_{i,k}),\varphi(p_{k,j})))
\end{eqnarray}

This isomorphism generalizes the concept of distance in weighted graphs. Using different TD-Norms, TD-Conorms we can calculate different types of distances and shortest paths in weighted graphs, such as: metric distances, ultra-metric distances, diffusion distances among an infinity of possibilities.

\begin{figure}[h]
\centering
\includegraphics[width=75mm,height=55mm]{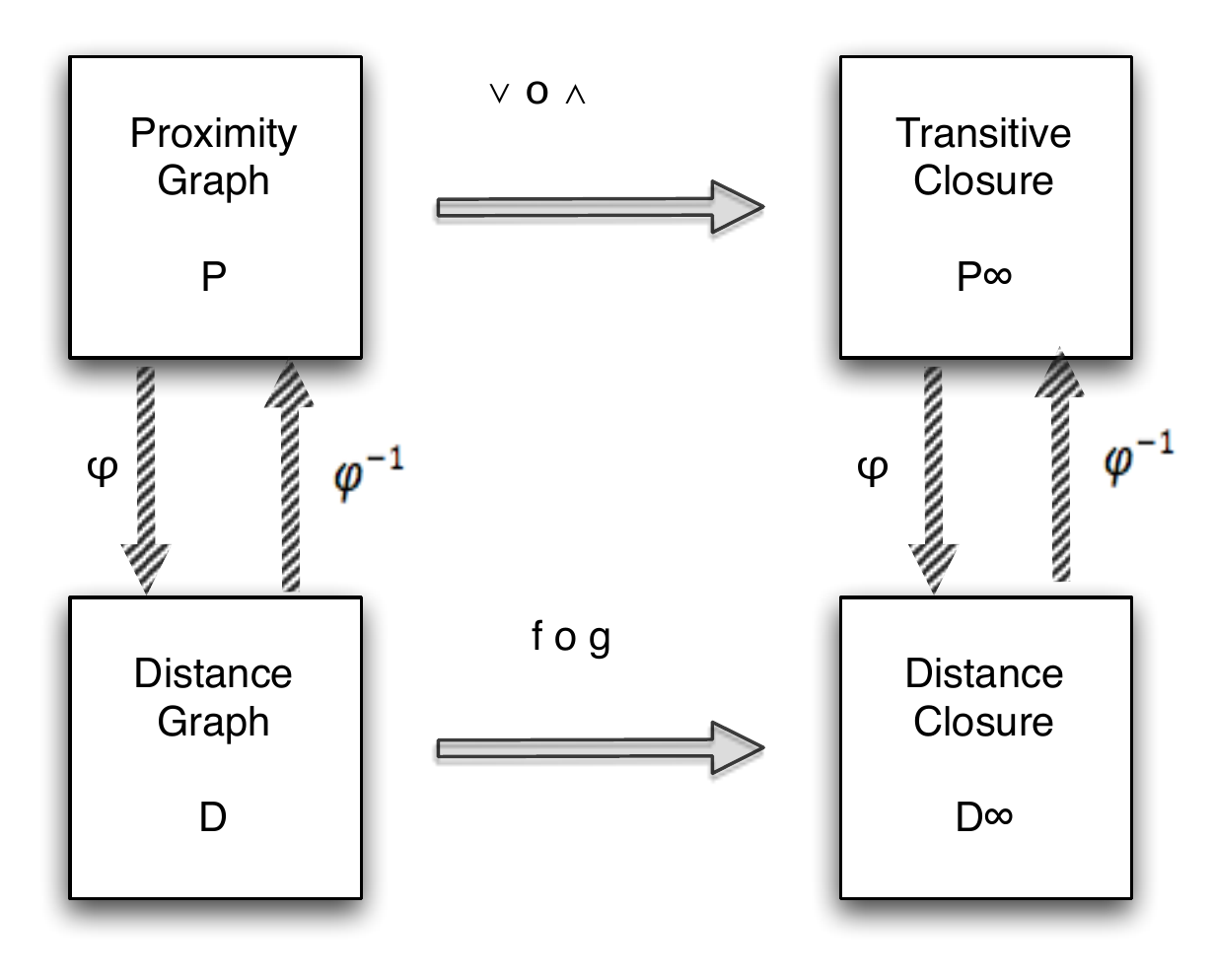}
\caption{Isomorphism between the proximity and distance spaces, with their respective transitive and distance Closures.\label{fig3}}
\end{figure}

\subsection{Semi-metric behavior}
A high value of proximity means that two items from one set (e.g.
words) tend to co-occur frequently in another set of objects (e.g.
web pages). But what about items that do not co-occur frequently
with one another, but do occur frequently with the same \emph{other}
elements? In other words, even if two items do not co-occur much,
they may occur very frequently with a third item (or more). Should
we infer that the two items are related via indirect associations,
that is, from \emph{transitivity}?
We would expect items that are strongly indirectly related to be more relevant than those that are not.

To build up a more intuitive understanding of transitivity in
weighted graphs, we convert our proximity graphs to distance graphs via isomorphism $\varphi$.
The simplest proximity-to-distance conversion
function is;

\begin{equation}
\label{phi_map_eq}
\label{distance_measures4} \varphi: d_{i,j}= \frac{1}{p_{i,j}} -
1
\end{equation}

A distance graph $D$, obtained via $\varphi$ from $P$ which is itself obtained from co-occurrence data in some corpus (as graphs $XYP$ and $YXP$), does not, in general, yield an Euclidean topology. This is because, for a pair of
elements $i$ and $j$, the triangle inequality
may be violated: $d_{i,j} \geq d_{i,k} + d_{k,j}$ for
some element $k$.
This means that the shortest distance between
two elements  may not be the direct edge but rather
an indirect path.
Distance functions that violate the triangle
inequality are referred to as \emph{semi-metrics}
\cite{galvin_shore91}.

Clearly, semi-metric behavior is a question of degree.
For some pairs of vertices in a distance graph an indirect path may
provide a much shorter indirect short-cut, a shorter distance, than
for others. To measure a degree of semi-metric behavior we have
introduced the \emph{semi-metric} and \emph{below average ratios}
\cite{Rocha2002}:

\[s_{i,j}=\frac{d_{i,j}}{\underline{d}_{i,j}}; \quad b_{i,j}=\frac{\overline{d_{i}}}{\underline{d}_{i,j}}
\]

\noindent where $\underline{d}_{i,j}$ is the shortest, direct or
indirect, distance between $i$ and $j$ in distance graph $D$,
and $\overline{d_{i}}$ is the mean direct distance from $i$ to
all other $k \in D$ such that $d_{i,k} \geq 0$.
$s_{i,j}$ is
positive and $> 1$ for semi-metric edges. $s_{i,j}$ and $b_{i,j}$ are only applied to
semi-metric edges $d_{i,j}$ where $0 <
\underline{d}_{i,j} < d_{i,j}$.
$b$ measures how much
the shortest indirect distance between $i$ and $j$ falls below
the average distance of $i$ to all its directly associated
elements $k$. The below average ratio is designed to capture
semi-metric behavior of non-finite edges: $d_{i,j} \rightarrow \infty$.
Note that $b_{i, j} \neq
b_{j,i}$. $b > 1$ denotes a below average distance reduction
(see \cite{Rocha2002} for more details).

\section{Recommendation from Proximity Graphs}
We developed and tested two types of collaborative filtering algorithms: proximity- and semi-metric-based. %Collaborative filtering systems start with the relation between Items and Users. This relation consists on the history or assessment done by the users to items. Examples are: the relation between users and items such as in Amazon.com where users buy books and other items, MovieLens where users rate movies, etc.
The training set is a relation between users ($U$) and items ($I$) from the past $R: U \times I$, where $r_{i,j} = 1$ if user $i$ has accessed item $j$, and $r_{i,j} = 0$ otherwise. This relation is a rectangular matrix of $n \times m$ entries.
Given $R$, using eq. \ref{proximity_measures}, we obtain user-based ($UIP$) and item-based ($IUP$) proximity graphs, as well as their isomorphic distance graphs obtained via the map of eq. \ref{phi_map_eq}. $UIP$ ($IUP$) is a weighted graph of $n$ ($m$) elements.
%
%We build our item-based  and user-based proximity graph by applying the generalized Jaccard similarity measure (see above).  Other measures can be used such as: cosine projection (vector based), mutual information, Pearson correlation and many more.
%
%Given the proximity between users (user-based) or the proximity between items (item-based) a collaborative filtering system identifies, for a given user, a set of items to be recommended. In the case of user-based approaches we search for a given user a neighborhood (using e.g. nearest neighbors method, see \cite{Fouss2}) of similar users and recommend the items more popular among the set of neighbors. In the case of item-based recommendation, for each user, we search for items similar to the ones that user have already consumed or rated.
%
%In this work, to show the evidence between semi-metric and causality we use the item-based approach.
%
%\subsection{Item-Based collaborative filtering recommendation system}
%Item-based systems are based on the proximities between items. Each user has associated a set of items, which is a subset of all items available to the users. Given the relation between items, we compute for each user a set of items which are similar to the items associated to that user.
%
Let us now describe our recommender algorithms based on these graphs:

\begin{Algorithm}{Item-Based Proximity}
\label{item-based}
\\
\emph{For each user $i=1\cdots n$:}
\begin{enumerate}
\item Retrieve the user vector $U_i$, containing the associated set of items from the training set $R$.

\item From $IUP$ remove all columns associated with items $j$ such that $r_{i,j}=0$ (items that do not appear in the user's profile from step $1$).

\item Calculate the mean value of row weights for each row in the reduced $IUP$ matrix obtained in step $2$. This results in a scalar score (in $[0,1]$) for all items $j=1\cdots m$.

\item User $i$ is recommended the top $n$ scored items.
\end{enumerate}
\end{Algorithm}

\begin{Algorithm}{Item-Based Semi-metric}
\label{item-based-SM}
Same as Algorithm \ref{item-based}, except that $IUP$ is enhanced with additional edges.
We calculate the metric closure from the proximity relation $IUP$ using the isomorphism of equation \ref{distance_measures4}. From the resulting distance graph, we identify the semi-metric pairs (edges) with below average ratio $b_{i,j}$  above a given threshold, and insert the corresponding edges from the transitive closure of $IUP^\infty$ into the original proximity graph ($IUP$). Finally we use this proximity graph as input for item-based proximity algorithm \ref{item-based}.
Notice that $IUP^\infty$ is, in this case, the isomorphic transitive closure to the metric closure of the distance graph. Therefore, the respective conjunction and disjunction operations employed are obtained from eq. \ref{isomorphism_formulae} for $f=min$ and $g=+$, given the isomorphism of eq. \ref{phi_map_eq}.
This results in $\vee = max$ and $\wedge = ab/(a+b -ab)$ (Hamacher product).
\end{Algorithm}

%\subsubsection{User-Based collaborative filtering recommendation system}

\begin{Algorithm}{User-Based Proximity}
\label{user-based}
\\
\emph{For each user $i=1\cdots n$:}
\begin{enumerate}
\item Determine the $k$ nearest users to user $i$ from proximity graph $UIP$: the $k$ highest values of row $i$ (neighborhood of user $i$ in graph $UIP$). % according to a given alpha-cut (threshold);
\item Recommend top $n$ most frequent items among neighborhood of user $i$ obtained in step $1$.
\end{enumerate}
\end{Algorithm}

%We tested the following user-based algorithms:

%\begin{enumerate}
%\item \emph{Proximity User-based algorithm (Prox-User-based)}\\
%The simplest user-based algorithm follows algorithm \ref{user-based} using the matrix $U\times U$ calculated from the proximity measure.% of equation \ref{proximity_measures}.
\begin{Algorithm}{User-Based Semi-metric}
\label{User-based-SM}\\
Here we enhance user proximity $UIP$ with semi.metric edges, just like we did for $IUP$ in algorithm \ref{item-based-SM}.
Afterwards, we use algorithm \ref{user-based}.
\end{Algorithm}

For both semi-metric algorithms (\ref{item-based-SM} and \ref{User-based-SM}), the thresholds for the below average ratio were set
on the distribution of $b_{i,j}$ around the cut-off point of the power law.

\section{Experimental Evaluation}
\subsubsection{Data Sets}
We used the benchmark data set of \emph{MovieLens}. This data set is a collection of votes, on a scale from one to five, given by web users (943 users) in respect to a given movie (1682 movies), as a total of 100,000 ratings. %The group Movilens provides a set of datasets. In these datasets were retained only users that had rated 20 or more movies (943 users). Each user gives his opinion (vote) in respect to a movie graded in a scale from one to five. These data sets are based in the full matrix items (movies) versus userÕs votes and partitioned in sets of training and test. It was studied by the group \cite{sarwar1} that the best partition is a training set with $80\%$ of the votes and a test set with the remaining $20\%$. These data sets are divided in two major data sets one with a test set with the ten votes per user (about 10,000 ratings), while the training set  contains the remainder of the ratings (about 90,000 ratings). None of the edges belongs both to the training and test sets.
In our experiment, to ascertain the utility of semi-metric behavior to predict user behavior, we do not need to use ratings; the goal is to predict which (future) movies, users will rate based on past behavior.  %, i.e., based on the past watched movies from a given user what are the movies he/she will watch in the future.
Therefore,  we converted ratings to binary votes: one (rated) or zero (not-rated).

\subsubsection{Evaluation Metrics}
We used the balanced $F1$ score, based on \emph{precision} and \emph{recall} measures, as well as variant of the \emph{Somers'D}, the degree of agreement
metric \cite{siegel}. Precision, recall, and the $F1$ measures are traditional measures in information retrieval, computed for unranked retrieval.
There are  other assessment measures for ranked results, as the Area Under the Precision and Recall Curve.
But since we compare our results to a previous benchmark effort that used the Somers'D measure on a
set of recommender systems \cite{fouss} \cite{Fouss2}, we also use it here.
Below, the measures employed are defined:

\begin{equation}
\label{recall}
recall=\frac{\mid test \cap top_n \mid}{\mid test \mid}
\end{equation}

\begin{equation}
\label{precision}
precision=\frac{\mid test \cap top_n \mid}{\mid top_n \mid}
\end{equation}

\begin{equation}
\label{f1}
F1=\frac{2 \cdot recall \cdot precision}{recall + precision}
\end{equation}

\noindent where $top_n$ is the set of top $n$ recommendations issued by a recommender system, and $test$ is the set of relevant or
expected recommendations from test set.
The variant of Somers'D method used for the MovieLens dataset, follows the following procedure described in \cite{Fouss2}.

\begin{enumerate}
\item For each user we take the row vector of similarities, $R:U\times I$, for each movie for the considered user.
%
%LMR: What is the "vector of similarities from training set" ? Not clear at all
%
\item Take only the non-watched movies for this user.
%
%LMR: you mean non-watched in training? I don't understand... Why? The Somers'D measure as explained here is not very clear.
%How do you use the test set to obtain a ranked set to compute formula below, if you are using only movies not in test set?
%
\item Rank the non-watched movies taking in consideration all movies.
\item Compute the degree of agreement: consider each pair $(a,b)$ of movies from recommended ranking, with $a$ in the test set and $b$ not.
If $a$ ahead of $b$: correct pair (agreement), $b$ ahead of $a$: incorrect pair, \ref{somers}.

\begin{equation}
\label{somers}
d=\frac{\# agreements}{\# total-of-pairs}
\end{equation}

\item Compute the global degree of agreement.

\end{enumerate}

%\subsubsection{Brief discussion on the two assess measures}
%Precision and Recall is a measure that have been largely used for evaluation of recommender systems. PR tries to capture a compromise between precision with recall of results. In our work we have considered the relevant set the set of all votes in the test set and the retrieved set the Top-N results obtained by the recommender system. Therefore recall is the ratio between the relevant results in the retrieved set and the relevant set. Precision is the ratio between the relevant results in the retrieved set and the number of results we want to retrieve, N.
%In general a user that uses a recommendation system only pays attention to the top N recommendation, i.e. the first page of recommendation, which is around the 10 first recommendations. In some recommendation systems like Amazon.com, perhaps a user can go a little bit further, but in a case of a Search Engines the user probably doesnÕt go much further.  Taking this in consideration a good recommender system is the one that has a high precision, even if the recall is not so good. The traditional F1 parameter gives the same importance to both precision and recall. Although, we think this way of assess a recommendation system is not the best because do not take into consideration the reality of the behavior of a user. However, it is common use on evaluation of recommender systems. Instead of using other Fn parameter, which gives for example more importance to precision, we prefer to use a variant of the SomersÕD degree of agreement.
This variant of Somers'D degree of agreement gives us a measure of how well our set of recommendations is distributed in the first positions of our list of relevant items. %This seems to capture better the performance of a recommender system \cite{Fouss2}.

%%%% AQUI

\section{Results}
%In this work we have set a experiment using the data sets described before. In the experiment we calculated the TFIDF (term frequency times Inverse Term Frequency), where by term we meant item and by document we meant user. In this way the movies that were seen by most of the users are given a less importance for the discrimination between users. We obtain better results with TFIDF compared with the original relation (vote matrix users by items); therefore in our experiment we used Fuzzy relation to assess our recommender systems.
%Our experiment consist by calculating the Jaccard proximity or cosine proximity from the TFIDF relation and then test it with the top 10 best recommendations. We have used several thresholds for the proximity aggregation with the goal to improve the similarities between users or items; however after using a range of different thresholds we obtain the same results for all algorithms.

%In the user-based collaborative filtering algorithms we have set parameters like number of neighbors and dimensionality reduction to the values suggested by Sarwar et al.  \cite{sarwar1} \cite{sarwar2}. We set the number of neighbors to be 90 and the dimensionality reduction to be 30 (the first 30 eigenvalues from the SVD decomposition).
We compare our results with the ones of Fouss et al \cite{Fouss2}. Table \ref{tableRS} shows our results for the proximity and semi-metric (SM) approaches for item- and user-based recommender systems.
Tables \ref{tableFouss} and \ref{tableFouss2}show the results obtained by Fouss et al in \cite{Fouss2} for several item- and user-based
recommender algorithms, respectively. %A more detailed description of these algorithms can be found in \cite{Fouss2}.
A good description of the algorithms involved in this comparison can be found in Fouss \cite{Fouss2}.
$L{+}$ is based on the pseudo-inverse of the Laplacian matrix; PCA CT is based on the principal component analysis of $L^{+}$;
kNN is based on the $k$-nearest neighbors algorithm;
Cosine is based on cosine similarity; Katz is based on the similarity index, which has been proposed in the social sciences field;
and Dijkstra based on the shortest paths of elements of the dataset.

\begin{table}[!th]
  \centering
   \scalebox{0.75}{
  \begin{tabular}{|c|c|c|c|c|} \hline
   & {Prox-Item-based} & {SM-Item-based}  & {Prox-User-based} & {SM-User-based} \\ \hline
   {Agreement (in \%)}& $89.53$ & $90.16$ & $88.20$ & $88.16$\\ \hline
   {F1}& $0.1827$ & $0.1832$ & $0.2130$ & $0.2179$\\ \hline
  \end{tabular}
  }
  \caption{Results for recommendation system. Somers'D degree of agreement \cite{fouss} \cite{Fouss2} and F1 measure.}
  \label{tableRS}
\end{table}

\begin{table}[!th]
  \centering
  \scalebox{0.8}{
  \begin{tabular}{|c|c|c|c|c|c|c|} \hline
   & {PCA CT}  & {$L^+$} & {kNN} & {Cosine} & {Katz} & {Dijkstra} \\ \hline
   {Agreement (in \%)}& $87.08$ & $90.99$ &$--$ & $--$ & $87.90$ & $49.11$\\ \hline
  \end{tabular}
  }
  \caption{Results for item-based recommendation systems from \cite{Fouss2}. }
  \label{tableFouss}
\end{table}

\begin{table}[!th]
  \centering
   \scalebox{0.8}{
  \begin{tabular}{|c|c|c|c|c|c|c|} \hline
   & {PCA CT}  & {$L^+$} & {kNN} & {Cosine} & {Katz} & {Dijkstra} \\ \hline
   {Agreement (in \%)}& $82.46$ & $93.02$ &$92.63$ & $92.73$ & $89.82$ & $76.09$\\ \hline
   {\#Neighbors}& $60$ & $100$ &$100$ & $60$ & $20$ & $100$\\ \hline
  \end{tabular}
  }
  \caption{Results for user-based recommendation system from \cite{Fouss2}.}
  \label{tableFouss2}
\end{table}

The semi-metric approach improves the item-based proximity method, in both $F1$ and the Somers'D measures (Table \ref{tableRS}), and is as good as the best item-based result reported in Fouss et al \cite{Fouss2} (Table \ref{tableFouss}).
Notice that performance measures (on a fixed gold standard) are not statistical, so all improvements are significant.
Our user-based algorithms are among the top such algorithms (table \ref{tableFouss2})---which tend to perform better than item-based algorithms, table \ref{tableFouss}, though in our approach the reverse was observed (Table \ref{tableRS}).
On our user-based approach, we see a slight improvement of including semi-metric edges with the $F1$ measure, but not with the Somers'D.
A possible explanation is the fact that user-based approaches depend on the number of neighbors around a given user. We leave an analysis of the impact of number of neighbors on our user-based method for future work, since the objective of this paper is simply to show that semi-metric behavior can improve recommender predictions.

\section{Discussion and Conclusions}
We show that exploring the natural clustering of proximity graphs (equations 1), leads to very simple, but competitive
item- and user-based recommender systems, in comparison to previous benchmarks in the literature \cite{Fouss2}.
Enhancing proximity graphs with semi-metric edges further improves recommendations, confirming the previous evidence in Rocha et al \cite{rocha05mylib}; on the item-based approach we see an improvement
in both $F1$ and Somers'D measures, while on the user-based approach we see it only on the $F1$ measure.
This improvement is not dramatic, but shows that semi-metric edges can be used to enhance prediction in recommender systems.
Since we barely scratched the surface of understanding semi-metric behavior in complex networks, the approach
is promising leaving plenty of room to improve the basic algorithms we introduced here.
%

%----------------------------------------------------------------------
\bibliographystyle{IEEEtran}
\bibliography{gbib}
\end{document}